\documentclass[twocolumn]{aastex61}
\received{July 1, 2016}
\revised{September 27, 2016}
\accepted{\today}
\submitjournal{ApJ}

\shorttitle{Two New ``Turn-off'' Changing-look  AGNs}
\shortauthors{Wang et al.}

\begin{document}

\title{Two New ``Turn-off'' Changing-look Active Galactic nuclei and Implication on ``Partially Obscured'' AGNs}

\correspondingauthor{J. Wang}
\email{wj@bao.ac.cn}

\correspondingauthor{D. W. Xu}
\email{dwxu@bao.ac.cn}

\author{J. Wang}
\affil{Guangxi Key Laboratory for Relativistic Astrophysics, School
of Physical Science and Technology, Guangxi University, Nanning
530004, People's Republic of China}
 \affil{Key Laboratory of Space
Astronomy and Technology, National Astronomical Observatories,
Chinese Academy of Sciences, Beijing 100101, China}

\author{D. W. Xu}
\affiliation{Key Laboratory of Space Astronomy and Technology, National Astronomical Observatories, Chinese Academy of Sciences, Beijing
100101, China}
\affiliation{School of Astronomy and Space Science, University of Chinese Academy of Sciences, Beijing, China}

\author{Y. Wang}
\affiliation{Key Laboratory of Optical Astronomy, National Astronomical Observatories, Chinese Academy of Sciences, Beijing 100101, China
}

\author{J. B. Zhang}
\affiliation{Key Laboratory of Optical Astronomy, National Astronomical Observatories, Chinese Academy of Sciences, Beijing 100101, China
}

\author{J. Zheng}
\affiliation{Key Laboratory of Optical Astronomy, National Astronomical Observatories, Chinese Academy of Sciences, Beijing 100101, China
}

\author{J. Y. Wei}
\affiliation{Key Laboratory of Space Astronomy and Technology, National Astronomical Observatories, Chinese Academy of Sciences, Beijing
100101, China}
\affiliation{School of Astronomy and Space Science, University of Chinese Academy of Sciences, Beijing, China}



\begin{abstract}

We here report a spectroscopic identification of two new changing-look AGNs (CL-AGNs): SDSS\,J104705.16+544405.8 
and SDSS\,J120447.91+170256.8 both with a ``turn-off'' type transition from type 1 to type 1.8/1.9.
The identification is arrived by a 
follow-up spectroscopic observation of the five changing-look AGN (CL-AGN) candidates that are 
extracted from the sample recently released in Macleod et al. The candidates are extract by the authors from the Sloan Digit 
Sky Survey Data 
Release 7 spectroscopically confirmed quasars with large amplitude variability.  
By compiling a sample of 26 previously identified CL-AGNs, we confirm the claim in Macleod et al. that CL-AGNs tend to be biased against 
low Eddington ratio, and identify an overlap between the CL-AGNs at their dim state and the so-called intermediate-type AGNs. 
The overlap implies that there two populations of the intermediate-type AGNs with different origins. One is due to the torus 
orientation effect, and the another the intrinsic change of the accretion rate of the central supermassive blackholes.
\end{abstract}




\section{Introduction} \label{sec:intro}

As a challenge to the widely accepted picture of active galactic nuclei (AGNs, e.g., Antonucci 1993; 
Shen \& Ho 2014),
the so-called ``changing-look'' (CL) phenomenon is a hot topic in modern astronomy.
Based on the generally accepted unified model (see a review in Antonucci 1993),
the observed type-I and -II spectra are explained by the orientation effect caused by the dust torus.
Type-I AGNs with both broad ($\mathrm{FWHM>1000\ km\ s^{-1}}$) and    
narrow ($\mathrm{FWHM\sim 10^2\ km\ s^{-1}}$) Balmer emission lines are believed to be observed face-on, 
while Type-II AGNs without the broad Balmer emission lines are observed edge-on.
In the CL phenomenon, an AGN changes its optical spectral type on a timescale of the
order of years, although Trakhtenbrot et al. (2019) recently claimed an identification of a CL phenomenon on 
a time scale of months in 1ES\,1927+654 by their high-cadence spectroscopic monitoring.

Up to date, there are only $\sim$70 CL-AGNs discovered by repeat and sparse spectroscopic observations, 
although both turn-on and turn-off transitions have been identified (e.g., Shapovalova et  
al. 2010; Shappee et al. 2014a; LaMassa et al. 2015; McElroy et al. 2016; Runnoe et al. 2016; Gezari et al.  
2017; Yang et al. 2018; Ruan et al. 2016; MacLeod et al. 2016, 2019; Wang et al. 2018; Stern et al. 2018). 
By reporting six turn-on CL-AGNs discovered during the first nine months operation of the 
Zwicky Transient Facilicty (ZTF) survey, Frederick et al. (2019) recently proposed a new class of 
CL-LINERs whose spectra in the quiescent state can be classified as a LINER with weak emission lines. 
An additional case of CL-LINER with a rapid ``turn-on'' transition to type-1 AGN was recently discovered in
SDSS\,J111536.57+054449.7 by Yan et al. (2019). 

In contrast to the CL phenomenon discovered in X-ray (e.g., Risaliti et al. 2009) which is generally 
believed to be caused by a large change of the line-of-sight absorption column (e.g., Matt et al. 2003; Piconcelli et al. 2007; 
Ricci et al. 2016), there is accumulating evidence supporting 
that the optical CL phenomenon is likely due to a variation in accretion  
rate of a supermassive blackhole (SMBH). The variation is resulted from either a viscous radial inflow or  
disk instability (e.g., Sheng et al. 2017, Yang et al. 2018; Wang et al. 2018;  
Gezari et al. 2017), although an explanation of accelerating outflow launched from the central SMBH
can not be entirely excluded (e.g., Shapovalova et al. 2010).
In the instability scenario, a thermal instability with a shorter timescale in optical region 
is involved additionally to solve the timescale problem (e.g., Lawrence 2018; Husemann et al. 2016).
Macleod et al. (2019) recently proposed that the optical CL phenomenon can be explained by an 
accretion supported broad-line region (BLR) in the disk-wind model (e.g., Nicastro 2000; Elitzur \& Ho 2009; Elitzur et al. 2014),
in which the BLR appears or disappears when the luminosity is above or below the critical one.

In this paper, we report an identification of two new CL-AGNs with a turn-off transition by repeat spectroscopy
of the  CL-AGN candidates recently selected by Macleod et al. (2019).
The paper is organized as follows. Section 2 describes the used sample. The observations and 
spectral analysis are presented in Section 3 and 4, respectively. A conclusion and discussion focusing on the 
dim state of a sample of CL-AGNs are 
presented in the last section. A $\Lambda$CDM cosmology with parameters
$H_0=70\mathrm{km\ s^{-1}\ Mpc^{-1}}$, $\Omega_m=0.3$, and
$\Omega_\Lambda=0.7$ is adopted throughout the paper.

\section{Sample} \label{sec:style}

Macleod et al. (2019) recently identified 
17 new CL-quasars (CLQs) by a follow-up spectroscopy to the highly variable SDSS DR7 spectroscopically identified quasars.
The variability is required to be $|\Delta g|>1$mag and $|\Delta r|>0.5$mag by 
comparing the photometric measurements between SDSS DR10 and Pan-STARRS (PS1, Kaiser et al. 2002).  
With the spectroscopic identifications, the authors claimed a CLQ confirmation rate of $\geq20\%$.
A catalog of more than 200 highly variable quasars was additionally released in Macleod et al. (2019)
for future follow-up spectroscopic identifications of new CL-AGNs. In order to ensure the H$\beta$ emission line in observer frame 
is within optical wavelength region, the redshifts of the candidates are limited to be smaller than 0.83.  

We performed a follow-up spectroscopic observation program by the 2.16m telescope at Xinglong observatory 
on a sub-sample of the highly variable quasars catalog given by Macleod et al. (2019).
After taking into account of both celestial location and brightness of the candidates, there are in total 
only five candidates available for the telescope before July.

\section{Observations and Data Reduction} \label{subsec:tables}

The follow-up spectroscopic observations and data reductions of the five CLQ candidates are
described in this section.

\subsection{Observations}

Our spectroscopic observations were carried out by the 2.16m telescope (Fan et al. 2016) at Xinglong observatory of National 
Astronomical Observatories, Chinese Academy of Sciences (NAOC) in several runs. The long-slit spectra were obtained by 
the Beijing Faint Object Spectrograph and Camera
(BFOSC) equipped with a back-illuminated E2V55-30 AIMO CCD as the detector. The grating G4 and a slit of width 1.8\arcsec\
oriented in the south-north direction were used in all the observation runs. This setup finally results in a spectral
resolution of $\sim$10\AA, as measured from the sky emission lines and comparison arcs, and provides a wavelength coverage from
3850 to 8000 \AA. Each target was observed either twice or triple in succession in each observation run. 
The exposure time of each frame ranges from 1200 to 2400 s. In each run, the wavelength calibration and
flux calibration were carried out by the iron-argon comparison arcs and by the Kitt Peak National Observatory (KPNO)
standard stars (Massey et al. 1988), respectively.
All the spectra were obtained as close to meridian as possible. 
The spectra of the standard stars close to the objects were observed with the same instrumental
setups.
Table 1 lists the log of observations of the five candidates, where Column (5) lists 
the total exposure in each run.

\begin{table}[h!]
\renewcommand{\thetable}{\arabic{table}}
\centering
\caption{Log of Spectroscopic Observation}
\label{tab:decimal}
\begin{tabular}{ccccc}
\tablewidth{0pt}
\hline
\hline
SDSS ID & $z$ & $g$-band & Date & Exposure \\
        &   &  mag   &      &  seconds \\
 (1)  & (2) & (3) & (4) & (5) \\
\hline
J085259.22+031320.6 & 0.297 & 16.19 & March 30 & 2400 \\
                    &       &       & March 31 & 2400 \\
J094443.08+580953.2 & 0.562 & 17.90 & March 23 & 2400 \\
                    &       &       & March 24 & 2400 \\
                    &       &       & March 25 & 2400 \\                    
J104705.16+544405.8 & 0.215 & 17.56 & April 09 & 2400 \\
                    &       &       & April 14 & 4800 \\
J105125.58+105621.5 & 0.602 & 18.07 & April 21 & 7200 \\
J120447.91+170256.8 & 0.298 & 16.69 & March 29 & 2400 \\
                    &       &       & April 03 & 4800 \\
                    &       &       & April 07 & 2400 \\
                    &       &       & April 13 & 4800 \\                    

\hline
\hline
\end{tabular}
\end{table}

\subsection{Data reduction}

We reduced the the 2D spectra in standard procedures by using the IRAF package\footnote{IRAF is distributed by the National Optical Astronomical Observatories,
which are operated by the Association of Universities for Research in
Astronomy, Inc., under cooperative agreement with the National Science
Foundation.}. The data reduction includes bias
subtraction, flat-field correction. The frames of each candidate obtained in the same night are 
combined to remove the contamination caused by cosmic-rays before the extraction of the 1D
spectrum. All the extracted 1D spectra were then calibrated in wavelength and flux by the corresponding
comparison arc and standards. The accuracy of the wavelength calibration is $\sim1$\AA. 
For each object, the calibrated spectra taken in different nights are 
combined to enhance the signal-to-noise ratio.

The Galactic extinction was corrected for each of the candidates by the extinction magnitude in $V$-band (Schlafly \& Finkbeiner 2011) 
taken from the NASA/IAPC Extragalactic Database (NED), assuming the $R_V=3.1$ extinction law of our Galaxy 
(Cardelli et al. 1989). Each of the spectra were then transformed to the rest frame, along with the correction of
the relativity effect on the flux, according to the corresponding redshift given by the SDSS pipelines, 
The rest frame specific flux is $f_{\lambda_{\mathrm{rest}}}=f_{\lambda_{\mathrm{obs}}}(1+z)^3$, where $f_{\lambda_{\mathrm{obs}}}$ is 
the specific flux in the observer frame and $\lambda_{\mathrm{rest}}=\lambda_{\mathrm{obs}}/(1+z)$. 

\subsection{Identification of changing-look phenomenon}

With our follow-up spectroscopy, a CL phenomenon with a turn-off transition 
can be clearly identified in two quasars: 
SDSS\,J104705.16+544405.8 and SDSS\,J120447.91+170256.8 (hereafter SDSS\,J1047+5444 and SDSS\,J1204+1702 for short). 
Figure 1 and Figure 2 compare \rm the SDSS and Xinglong spectra for the 
two CLQs and the three non-CLQs, respectively. The signal-to-noise ratio of the Xinglong spectrum of 
SDSS\,J094443.08+580953.2 is too low to 
enable us to given any meaningful result on this object. \rm In the comparison, the spectra taken by SDSS are convolved with a Gaussian profile to match the 
spectral resolution of the Xinglong spectra. The flux level of each Xinglong spectrum is scaled by a factor determined by 
requiring the modeled total [\ion{O}{3}]$\lambda$5007 line flux equals to that of the corresponding SDSS spectrum (see Section 4). 

One can see clearly from the comparison a turn-off type transition from classical type-1 to type-1.9 in both quasars, i.e., SDSS\,J1047+5444 and SDSS\,J1204+1702. The H$\beta$ and H$\gamma$ broad components 
almost disappear in both Xinglong spectra, along with a significant weakening of both H$\alpha$ broad emission and
AGN's featureless continuum. In fact, by a direct integration over a proper wavelength range on the residual spectra, 
the relative variation of the H$\beta$ broad emission line $\Delta f/f_{\mathrm{SDSS}}$ is estimated to 
to be $\sim-0.72$ and $-0.67$ for SDSS\,J1047+5444 and SDSS\,J1204+1702, respectively, where $f_{\mathrm{SDSS}}$ is the line flux 
obtained from the SDSS spectra and $\Delta f=f_{\mathrm{Xionglong}}-f_{\mathrm{SDSS}}$. The value of $\Delta f/f_{\mathrm{SDSS}}$
is, however, as low as -0.01, 0.14, and 0.02 for the other three non-CLQs.
In SDSS\,J1047+5444, We argue that the continuum of the Xinglong spectrum is changed to be even dominated by the 
host stellar emission with a 4000\AA\ break due to the stellar matal absorptions and a marginally detected \ion{Mg}{1}b (5176\AA) 
absorption feature that is marked in Figure 1. \rm

\begin{figure}
\plotone{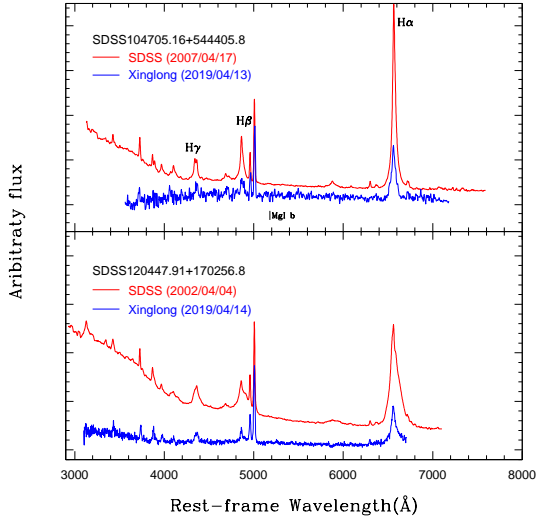} \caption{
Comparison between the Xinglong spectra and the spectra extracted from the SDSS DR7 archive database. 
The SDSS spectra are convolved with a Gaussian profile to give a spectral resolution being identical to that of the Xinglong spectra.
For each object, the two spectra are scaled to have a common flux of the total [\ion{O}{3}]$\lambda$5007 \rm line flux.
All the spectra are transformed to the rest-frame, and are shifted vertically by an arbitrary amount for visibility.
The fist three Balmer lines, H$\alpha$ (6563\AA), H$\beta$ (4861\AA) and H$\gamma$ (4340\AA), are labeled on the upper panel. 
}
\end{figure}

\begin{figure}
\plotone{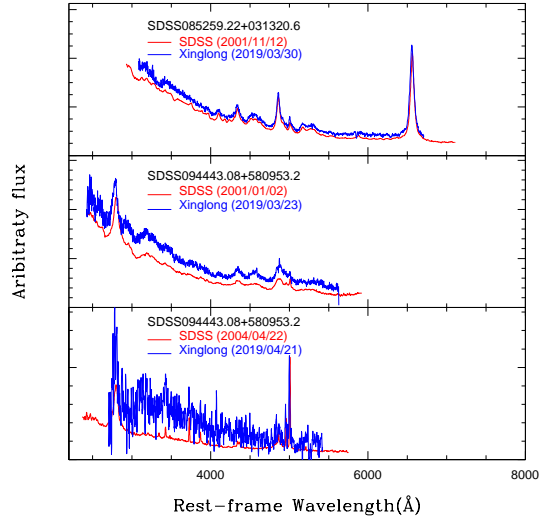} \caption{
The same as in Figure 1 but for the other three non-CLQs. 
}
\end{figure}

\section{Spectral Analysis}

A spectral analysis is performed by following Wang et al. (2018, and references therein) in this section 
to shed a light on the turn-off type transitions occurring in the two CLQs.

\subsection{AGN Continuum and Stellar Feature Removal}

We at first model the continuum of each of the four spectra by a linear combination 
of the following components: (1) an AGN's powerlaw continuum; (2) an template of both high-order Balmer emission
lines and a Balmer continuum from the BLR; (3) an template of both 
optical and ultraviolet \ion{Fe}{2} complex; (4) a host galaxy template with an age of 5Gyr extracted
from the single stellar population (SSP) spectral library given in Bruzual \& Charlot (2003);
(5) an intrinsic extinction due to the host galaxy described by a galactic extinction
curve with $R_V=3.1$. We use the empirical optical \ion{Fe}{2} template provided by Veron-Cetty et al. (2004) and
the theoretical template by Bruhweiler \& Verner (2008) to model the optical and ultraviolet \ion{Fe}{2} complex, respectively. 
The line width of the template is fixed in advance to be that of the broad component of H$\beta$, which is determined by our 
line profile modeling (see below).

The emission from a partially optically thick cloud with an electron temperature of $T_e=1.0\times10^4$K is adopted to 
model the Balmer continuum $f_\lambda^{\mathrm{BC}}$ by following Dietrich et al.
(2002, see also in Grandi 1982 and Malkan \& Sargent 1982):
\begin{equation}
  f_\lambda^{\mathrm{BC}}=f_\lambda^{\mathrm{BE}}B_\lambda(T_e)(1-e^{-\tau}),\  \lambda\leq\lambda_{\mathrm{BE}}
\end{equation}
where $f_\lambda^{\mathrm{BE}}$ is the continuum flux at the Balmer edge $\lambda_{\mathrm{BE}}=3646$\AA
and $B_\lambda(T_e)$ is the Planck function. $\tau_\lambda$ is the optical depth
at wavelength $\lambda$, which is related to the one at the Balmer edge $\tau_{\mathrm{BE}}$ as 
$\tau_\lambda=\tau_{\mathrm{BE}}(\lambda/\lambda_{\mathrm{BE}})^3$. A typical value of $\tau_{\mathrm{BE}}=0.5$ is
adopted in our modeling of the continuum.

We model the high-order Balmer lines (i.e., H7-H50) by the case B recombination model with an electron temperature of
$T_e=1.5\times10^4$K and an electron density of $n_e=10^{8-10}\ \mathrm{cm^{-3}}$ (Storey \& Hummer 1995). 
The widths of these high-order Balmer lines are, again, determined in advance according to the line profile
modeling of the H$\beta$ broad emission (see below).

A $\chi^2$ minimization is performed iteratively over the whole spectroscopic wavelength range, 
except for the regions with known emission lines (e.g., H$\alpha$, H$\beta$, H$\gamma$, H$\delta$, [\ion{S}{2}]$\lambda\lambda6716,6731$,
[\ion{N}{2}]$\lambda\lambda6548,6583$, [\ion{O}{1}]$\lambda$6300, [\ion{O}{3}]$\lambda\lambda4959,5007$, [\ion{O}{2}]$\lambda3727$,
[\ion{Ne}{3}]$\lambda3869$, and [\ion{Ne}{5}]$\lambda$3426). 
For both SDSS spectra being typical of a type-I AGN, the underlying stellar emission is failed to be modeled because the continuum is
entirely dominated by the AGN's featureless emission. The modeling of the underlying stellar emission is also failed in 
the Xinglong spectrum of SDSS\,J1204+1702 due to the poor S/N ratio of its continuum. In contrast, 
the Xinglong spectrum of SDSS\,J1047+5444 shows a continuum being typical of an intermediate-type AGN with weak or negligible emission 
from the central engine, although our modeling based on
a 5Gyr old SSP returns a poor reproduction of the continuum. In addition,  
our modeling suggests that the optical \ion{Fe}{2} complex is too weak to be modeled in all the four spectra.  
The removal of the modeled continuum is illustrated in Figure 3 for the four spectra.

\begin{figure}
\plotone{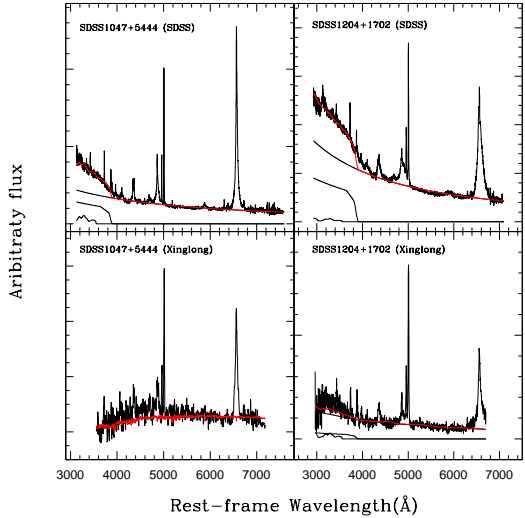} \caption{
Illustration of the modeling and removal of the continuum for the SDSS DR7 and Xinglong spectra. In each panel, the top, heavy 
curve shows the observed rest-frame spectrum overplotted by the best-fitted continuum shown by the red curve. The light curves underneath 
show the individual components used in the modeling. }
\end{figure}

\subsection{Line Profile Modeling}

For each emission-line-isolated spectrum, 
the emission line profiles are modeled on both H$\alpha$ and H$\beta$ regions by the SPECFIT task (Kriss 1994)
in the IRAF package. 

In the profile modeling, each emission line is reproduced by a set of Gaussian profiles.
By taking the poor continuum removal described above into account,
a local linear continuum is additionally used for the Xinglong spectrum of SDSS\,J1047+5444 to reproduce the line profiles 
in the H$\beta$ region. 
The line flux ratios of the [\ion{O}{3}]$\lambda\lambda$4959, 5007 and [\ion{N}{2}]$\lambda\lambda$6548, 6583
doublets are fixed to their theoretical values, i.e., 1:3. The line widths of both H$\alpha$ and H$\beta$ broad components are 
measured directly on the residual profiles by the SPLOT task in the IRAF package after subtracting the modeled narrow
components from the observed profiles. 
The line modelings are shown in the left and right panels of  Figure 4  for 
the H$\beta$ and H$\alpha$ regions, respectively. The results of the profile modeling are
listed in Table 2. 
No intrinsic extinction correction is applied to all the derived line fluxes, because the
Balmer decrements of the narrow components determined from the SDSS spectra are as small as 
$\mathrm{H\alpha/H\beta}=2.87\pm0.17$ and $1.43\pm0.14$ for SDSS\,J1047+5444 and SDSS\,J1204+1702, respectively. 
All the errors reported in the table correspond to the 1$\sigma$ significance level after taking
into account the proper error propagation.

\begin{figure}
\plotone{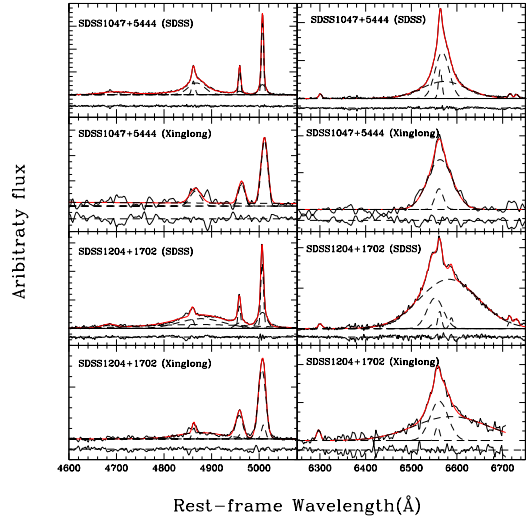} 
\caption{Line profile modelings by a linear combination of a set of Gaussian functions for the H$\beta$ (the left panels) and 
H$\alpha$ (the right panels) regions. In each panel the modeled continuum has already been removed from the original observed spectrum. 
The observed and modeled line profiles are plotted by black and red solid lines, respectively. Each Gaussian function is shown by a
dashed line. The sub-panel underneath the line spectrum presents the residuals between the observed and modeled profiles. 
}
\end{figure}

\begin{table*}[h!]
\tiny
\renewcommand{\thetable}{\arabic{table}}
\centering
\caption{Spectral measurements and analysis.} \label{tab:decimal}
\begin{tabular}{cccccccccc}
\tablewidth{0pt}
\hline
\hline
SDSS ID & Epoch & AGN type & $F(\mathrm{[OIII]\lambda5007})$ & $F(\mathrm{H\beta_b})$ & FWHM(H$\beta_\mathrm{b}$) & $F(\mathrm{H\alpha_b})$ & 
FWHM(H$\alpha_\mathrm{b}$) & $M_{\mathrm{BH}}/M_\odot$ & $L_{\mathrm{bol}}/L_{\mathrm{Edd}}$\\
& & & \multicolumn{2}{c}{$\mathrm{10^{-15}\ erg\ s^{-1}\ cm^{-2}}$}  & $\mathrm{km\ s^{-1}}$ & 
$\mathrm{10^{-15}\ erg\ s^{-1}\ cm^{-2}}$ & $\mathrm{km\ s^{-1}}$ & & \\
(1) & (2) & (3) & (4) & (5) & (6) & (7) & (8) & (9) & (10)\\
\hline
J1047+5444 & 2002/04/04 & 1.0     & $13.5\pm0.02$ &  $24.5\pm1.1$ & $2600\pm120$ & $102.5\pm1.1$  & $1970\pm80$ & $4.0\times10^7$ & 0.41 \\
               & 2019/04/14 & 1.9     & $9.1\pm0.2$   &  \dotfill     & \dotfill     & $18.7\pm0.4$   & $2940\pm60$ & $3.6\times10^7$ & 0.15 \\
J1204+1702 & 2007/04/17 & 1.0     & $15.2\pm0.3$  &  $40.0\pm1.8$ & $6720\pm200$ & $115.5\pm1.0$  & $4920\pm50$ & $3.7\times10^8$ & 0.09 \\
               & 2019/04/14 & 1.8/1.9 & $14.7\pm0.9$  &  $7.1\pm0.4$  & $6810\pm600$ & $35.7\pm1.0$   & $3140\pm130$ & $1.0\times10^8$ & 0.12 \\
\hline
\end{tabular}
\end{table*}

\subsection{Blackhole mass and Eddington ratio}

Based on the line profile modelings, the SMBH viral mass ($M_{\mathrm{BH}}$) and Eddington ratio $L_{\mathrm{bol}}/L_{\mathrm{Edd}}$ 
(where $L_{\mathrm{Edd}}=1.26\times10^{38}M_{\mathrm{BH}}/M_\odot\ \mathrm{erg\ s^{-1}}$ is the Eddington luminosity)
are estimated for the two CLQs from the single-epoch spectroscopy through several well-established calibrated relationships (e.g., 
Wu et al. 2004; Kaspi et al. 2000, 2005; Peterson \& Bentz 2006; Marziani \& Sulentic 2012; Du et al. 2014, 2015; 
Peterson 2014; Wang et al. 2014).       
 
Our estimations of both $M_{\mathrm{BH}}$ and $L_{\mathrm{bol}}/L_{\mathrm{Edd}}$ are based on the modeled broad H$\alpha$ emission lines. 
Green \& Ho (2005) provided a calibration of
\begin{equation}
  M_{\mathrm{BH}}=3.0\times10^6\bigg(\frac{L_{\mathrm{H\alpha}}}{10^{42}\ \mathrm{erg\ s^{-1}}}\bigg)^{0.45}
  \bigg(\frac{\mathrm{FWHM(H\alpha)}}{1000\ \mathrm{km\ s^{-1}}}\bigg)^{2.06}M_\odot
\end{equation}
to estimate $M_{\mathrm{BH}}$. To obtain an estimation of $L_{\mathrm{bol}}/L_{\mathrm{Edd}}$, we derive the bolometric luminosity
$L_{\mathrm{bol}}$ from the standard bolometric correction $L_{\mathrm{bol}}=9\lambda L_\lambda(5100\AA)$ (e.g., Kaspi et al. 2000),
where $L_\lambda(5100\AA)$ is the AGN's specific continuum luminosity at 5100\AA, which can be inferred from
the H$\alpha$ broad-line luminosity through the calibration (Greene \& Ho 2005)
\begin{equation}
 \lambda L_\lambda(5100\AA)=2.4\times10^{43}\bigg(\frac{L_{\mathrm{H\alpha}}}{10^{42}\ \mathrm{erg\ s^{-1}}}\bigg)^{0.86}
\end{equation}

The estimated $M_{\mathrm{BH}}$ and $L_{\mathrm{bol}}/L_{\mathrm{Edd}}$ are tabulated in the Column (9) and (10) in Table 2, respectively. 
For each object, the used H$\alpha$ line flux measured from the Xinglong spectrum is 
scaled by a factor determined by equaling the total [\ion{O}{3}]$\lambda5007$ line flux to that of the corresponding SDSS 
spectroscopy.

For SDSS\,J1047+5444, the two spectra taken at different epochs return consistent estimations of the $M_{\mathrm{BH}}$.
The corresponding $L/L_{\mathrm{Edd}}$ decreases from 0.41 to 0.15. For SDSS\,J1204+1702, we, however, obtain a roughly constant 
$L/L_{\mathrm{Edd}}$ when the object is at the ``on'' and ``off'' states. We argue that the invariable 
$L/L_{\mathrm{Edd}}$ is due to the difference in the estimated $M_{\mathrm{BH}}$, in which the $M_{\mathrm{BH}}$ determined 
from the SDSS spectrum is about four times larger than that from the Xinglong spectrum. By adopting the $M_{\mathrm{BH}}$ 
from the SDSS spectrum, similar as in SDSS\,J1047+5444, the corresponding $L/L_{\mathrm{Edd}}$ in fact decreases from 0.09 to 0.02.

\section{Conclusion and Discussion}

By performing a follow-up spectroscopy on five CL-AGN candidates recently selected by 
Macleod et al. (2019), we identify two new CL quasars, i.e., SDSS\,J1047+5444 and SDSS\,J1204+1702, both with a 
``turn-off'' type transition, when the new spectra taken by the 2.16m telescope in Xinglong observatory are compared to
the SDSS archival spectra. 

With the increasing number of the identified CL-AGNs, there is accumulating evidence supporting that 
the change of SMBH's accretion rate is the physical origin of the CL phenomenon (e.g., LaMass et al. 2015; 
Ruan et al. 2016; Runnoe et al. 2016; Gezari et al. 2017; Yang et al. 2018; Sheng et al. 2017; Wang et al. 2018; 
Yan et al. 2019; Macleod et al. 2016, 2019). 
The multi-wavelength light curves of the two newly identified CL-AGNs are plotted in Figure 5. 
Both objects show a continual decrease of the mid-infrared (MIR) brightness detected by the Wide-field Infrared Survey Explorer 
(WISE and NEOWISE-R, Wright et al. 2010; Mainzer et al. 2014) from 2010 to 2017, i.e., within the two spectroscopic epochs . The 
fading of the MIR emission supports the fact that the identified ``turn-off'' type transitions in the two 
objects are more likely due to the decrease of accretion rate rather than obscuring, because the 
MIR emission is mainly resulted from AGN-heated hot-dust which is less sensitive to dust obscuring 
(e.g., Sheng et al. 2017; Stern et al. 2018).
Macleod et al. (2019) recently argued that the CLQs are 
the extreme tail of regular quasar variability (e.g, Rumbaugh et al. 2018). The argument is based on the fact that 
the CLQs are found to have relatively low $L/L_{\mathrm{Edd}}$, which is consistent with the well-known anti-correlation 
between $L/L_{\mathrm{Edd}}$ and variability amplitude previously revealed in quasars (e.g., Wilhite et al. 2008; Mao et al. 2009; Macleod et al. 2010). 

\begin{figure}
\plotone{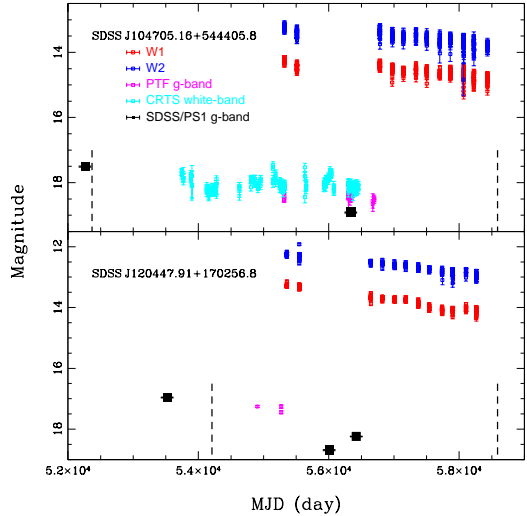} 
\caption{The multi-wavelength light curves for the two newly identified CL-AGNs. In 
each panel, the two vertical dashed lines mark the epochs of the SDSS and Xinglong spectra.}
\end{figure}

We compile a sample of 26 previously identified CL-AGNs from literature by requiring the detailed measurements
on their both on and off states are provided in literature\footnote{In the sample given in Yang et al. (2018), 
there are only 3 common objects with $z<0.5$ listed in the value-added SDSS DR7 quasar catalog published in Shen et al. (2011).  
The CL-AGN sample in Macleod et al. (2016) is not included in the current study because there was no spectral measurements for the 
repeat spectroscopies. The objects associated with a type transition between a quiescent LINERs and an AGN (Frederick et al. 2019; 
Yan et al. 2019) are not included in the sample.}.
Figure 6 shows the distributions of the CL-AGNs
on the $L_{\mathrm{bol}}$ versus $M_{\mathrm{BH}}$ (the left panel) and $L_{\mathrm{bol}}$ versus $L/L_{\mathrm{Edd}}$ 
(the right panel) diagrams.  The on and off states are plotted in the 
diagrams by the open and solid squares for each CL-AGN, respectively. The measurements given in Shen et al. (2011) and Chen et al. (2018) are 
adopted for the on state for each of the objects. For the off state, the value of $L_{\mathrm{bol}}$ is 
obtained from the corresponding value at the on state by a scaling factor determined through the change of the broad H$\alpha$ line flux. 
The comparison samples used in the diagrams include 1) the SDSS DR7 quasars with $z<0.5$ (Shen et al. 2011); 2) the 
SDSS DR3 narrow-line Seyfert 1 galaxies given in Zhou et al. (2006); 3) the \it Swift/\rm BAT AGN sample with a spectral type classification 
in Winter et al. (2012); and 4) the SDSS intermediate-type Seyfert galaxies studied in Wang (2015).

Two facts can be learned from the comparison shown in Figure 6.   On the one hand, as being consistent with Macleod et al. (2019), 
one can see from the figures that the CL-AGNs at on state are biased towards both low $L_{\mathrm{bol}}$ and 
low $L/L_{\mathrm{Edd}}$. This bias in fact motivates Macleod et al. (2019) to argue that the disk-wind BLR models proposed in 
Elitzur \& Ho (2012) and Nicastro (2000) are plausible for understanding the (dis)appearance of the broad emission lines observed in the 
CL phenomenon, although a critical value of $L/L_{\mathrm{Edd}}\sim10^{-3}$ is required for 
the (dis)appearance when the fiducal values of a set of parameters of the disk are adopted.   
On the other hand, there is an overlap between the intermediate-type AGNs and the CL-AGNs at their off state. By adopting the change of accretion rate as the physical origin of the CL phenomenon, this overlap strongly implies that the 
so-called intermediate-type AGNs are composed of two populations. One is due to the well-accepted orientation effect, and 
the another the intrinsic change of accretion rate, even though the physical origin of the change is still unclear at the 
current stage. In fact, the analysis of the X-ray spectra in Winter et al. (2012) suggests that the Seyfert-1.5 galaxies
statistically show higher neutral column densities than the Seyfert 1 galaxies, which agrees with the expectation of the unified model 
(e.g., Antonucci 1993).
The statistics in Figure 6 suggests that the two populations might differ from each other in $L_{\mathrm{bol}}$. 

The expected timescale is still an open issue in the scenario of change of SMBH's accretion rate (i.e., the viscosity crisis, e.g., 
Lawrence 2018 and references therein). 
The viscous timescale of a viscous radial inflow is expected to be (e.g., Krolik 1999; Shakura \& Sunyaev 1973; 
LaMassa et al. 2015; Gezari et al. 2017)
\begin{equation}
  \footnotesize
  t_{\mathrm{infl}}=6.5\bigg(\frac{\alpha}{0.1}\bigg)^{-1}\bigg(\frac{L/L_{\mathrm{Edd}}}{0.1}\bigg)^{-2}\bigg(\frac{\eta}{0.1}\bigg)^2
  \bigg(\frac{r}{10r_g}\bigg)^{7/2}\bigg(\frac{M_{\mathrm{BH}}}{1\times10^8M_\odot}\bigg)\ \mathrm{yr}
\end{equation}
where $\alpha$ is the ``viscosity'' parameter, $\eta$ is the efficiency of converting potential energy to radiation, and 
$r_g$ is the gravitational radius in unit of $GM/c^2$. Adopting $\alpha=\eta=L/L_{\mathrm{Edd}}=0.1$ and 
$M_{\mathrm{BH}}=1\times10^8M_\odot$ yields a viscous timescale being comparable to the observations for the near-ultraviolet
emission radiated from inner accretion disk with $r\sim10r_g$. The timescale, however, dramatically increase to $\sim10^3$yr  
for the optical emission coming from the outer disk with $r\sim50-100r_g$. 

Some solutions have been proposed to alleviate the timescale crisis. One solution is to involve the 
local disk thermal instability. The evolutionary $\alpha$-disk model developed in Siemiginowska et al. (1996) predicts 
a thermal timescale of 
\begin{equation}
 t_{\mathrm{th}}=2.7\bigg(\frac{\alpha}{0.1}\bigg)^{-1}\bigg(\frac{r}{10^{16}\ \mathrm{cm}}\bigg)^{3/2}\bigg(\frac{M_{\mathrm{BH}}}{10^8\ M_\odot}\bigg)^{1/2}\ \mathrm{yr} 
\end{equation}
Husemann et al. (2016) in fact reveals a temperature variation in Mrk\,1018. The simulation carried out by Jiang et al. (2016)
suggests that the development of the disk thermal instability favors low-metallicity gas.
An alternative solution is to involve an accretion disk elevated by a magnetic field, which can results in a shorter 
variability timescale and can explain the CL phenomenon by an abrupt variation in magnetic torque (e.g., Ross et al. 2018; Stern et al. 2018; 
Dexter \& Begelamn 2019). 

\begin{figure}
\plotone{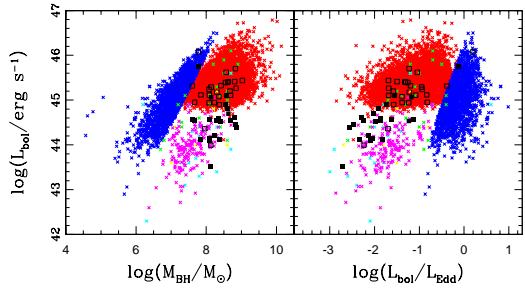} 
\caption{The distributions of the 26 previously identified CL-AGNs on the $L_{\mathrm{bol}}$-$M_{\mathrm{BH}}$ (the left panel) and 
$L_{\mathrm{bol}}$-$L_{\mathrm{bol}}/L_{\mathrm{Edd}}$ (the right panel) diagrams. The on and off states are 
denoted by the black open- and solid-squares, respectively. The used comparison samples are described as follows. 
Red cross: the quasars with $z<$0.5 taken from the value-added SDSS DR7 quasar catalog (Shen et al. 2011); blue cross: 
the SDSS DR3 NLS1 catalog established by Zhou et al. (2006); magenta cross: the SDSS DR7 intermediate-type AGNs studied in 
Wang (2015). The \it Swift\rm/BAT AGN sample in Winter et al. (2012) is shown by the green, yellow and cyan cross for 
Seyfert 1, 1.2 and 1.5 galaxies, respectively. }
\end{figure}

The CL phenomenon is far from being understood at the current stage partially because of the limited CL-AGN sample size.   
Both repeat imaging and spectroscopy are necessary for expanding the sample size. 
Although some ongoing and forthcoming optical survey programs (e.g., Pan-STARRS1
survey, Chambers et al. (2016), Zwicky Transient Facility, Kulkarni (2018), All-Sky Automated Survey
for Supernovae, Shappee et al. (2014), Large Synoptic Survey Telescope project, LSST
Science Collaboration et al. (2017)), can provide many interesting targets for 
follow-up spectroscopy, how to flag the CL phenomenon efficiently by excluding the contamination
caused by the AGN’s normal variation in optical bands is an open issue. This issue can be fairly addressed by some recently
proposed space-based ultraviolet (UV) patrol missions (e.g., Wang et al. 2019; Sagiv et al. 2014; Mathew
et al. 2018), because a significant variation in
the AGN's UV continuum is expected in the CL phenomenon. Additionally, a better understanding of the CL phenomenon
can stem from the forthcoming larger time-domain spectroscopic surveys, 
for example the SDSS-V survey (Kollmeier et al. 2017).

\acknowledgments

The authors thank the anonymous referee for his/her careful review and
helpful suggestions that improved the manuscript.
JW \& DWX are supported by the National Natural Science Foundation of China under grants
11773036. The study is supported by the National Basic Research Program of China (grant
2014CB845800), by Natural Science Foundation of Guangxi (2018GXNSFGA281007), 
Bagui Young Scholars Program, and by the Strategic Pioneer Program on Space Science, Chinese Academy of Sciences (Grant
No. XDA15052600 \& XDA15016500).
This work is partially supported by the Open Project Program of the Key Laboratory of Optical Astronomy, NAOC, CAS.
This study uses the SDSS archive data that was created and distributed by the Alfred P.
Sloan Foundation, the Participating Institutions, the National Science
Foundation, and the U.S. Department of Energy Office of Science.
Special thanks go to the staff at Xinglong Observatory for their instrumental and observational helps, 
and to the allocated observers who allowed us to finish the observations in ToO mode.
\vspace{5mm}
\facilities{Xinglong 2.16m telescope}
\software{IRAF (Tody 1986, Tody 1993)}

\end{document}